\documentclass[
reprint,  
amsmath,amssymb,
aps,
]{revtex4-2}
\usepackage[utf8]{inputenc}
\usepackage[version=3]{mhchem}
\usepackage{amsmath}
\usepackage{color}
\usepackage{titlesec}
\usepackage{soul}
\usepackage{braket}
\usepackage{xr-hyper}
\usepackage{hyperref}
\usepackage{graphicx}
\usepackage{dcolumn}
\usepackage{bm}
\usepackage{tikz}
\usepackage{ragged2e}
\usepackage{kantlipsum}
\usepackage{setspace}
\usepackage{upgreek}
\usepackage{capt-of}
\usepackage{float}
\usepackage{cancel}
\usepackage[normalem]{ulem}
\usepackage{lineno}   
\usepackage{comment}
\allowdisplaybreaks 
\usetikzlibrary{matrix}
\usepackage{eqparbox}
\newbox\matrixcellbox
\tikzset{center align per column/.style={nodes={execute at begin
			node={\setbox\matrixcellbox=\hbox\bgroup},
			execute at end
			node={\egroup\eqmakebox[\tikzmatrixname\the\pgfmatrixcurrentcolumn][c]{\copy\matrixcellbox}}}},
}

\titleformat{\section}{\centering\normalfont\bfseries}{\thesection}{1em}{}
\titlespacing{\section}{0pt}{*2}{*1}
\titlespacing{\subsection}{0pt}{*2}{*1}

\usepackage{xcolor}
\usepackage[draft,inline,nomargin,index]{fixme}

\fxsetup{theme=color,mode=multiuser}

\FXRegisterAuthor{ofer}{aOfer}{\color{brown} \bf Ofer} 
\FXRegisterAuthor{bankim}{abankim}{\color{red} \bf Bankim}
\FXRegisterAuthor{ashley}{aashley}{\color{magenta} \bf Ashley}
\FXRegisterAuthor{aditya}{aaditya}{\color{blue} \bf Aditya}
\FXRegisterAuthor{ariel}{aariel}{\color{orange} \bf Ariel}
\FXRegisterAuthor{general}{ageneral}{\color{violet} \bf Collective}
\definecolor{cadmiumgreen}{rgb}{0.0, 0.68, 0.24}

\newcommand{\rb}{r_\mathrm{b}}
\newcommand{\OD}{\mathrm{OD}}
\newcommand{\ODb}{\mathrm{OD}_\mathrm{b}}
\newcommand {\rmi}{{i}}

\newcommand{\gtwo}{g^{(2)}}
\newcommand{\gthree}{g^{(3)}}
\newcommand{\mus}{\mathrm{\mu s}}
\newcommand{\mum}{\mathrm{\mu m}}
\newcommand{\gammaEIT}{\gamma_\mathrm{E}}
\newcommand{\la}{l_\mathrm{a}}
\newcommand{\vg}{v_\mathrm{g}}

\newcommand{\tcross}{\tau_\mathrm{cross}}
\newcommand{\tself}{\tau_\mathrm{self}}
\begin{document}
	
	\title{Long-Range Blockade Between Counter-Propagating Photons}
	
	\author{Bankim Chandra Das$^{*}$}
	\affiliation{Department of Physics of Complex Systems, Weizmann Institute of Science, Rehovot, Israel
		\textnormal{($^*$Contributed equally)}}
	\author{Ashley Harkavi$^{*}$}
	\affiliation{Department of Physics of Complex Systems, Weizmann Institute of Science, Rehovot, Israel
		\textnormal{($^*$Contributed equally)}}
	\author{Aditya Prakash$^{*}$}
	\affiliation{Department of Physics of Complex Systems, Weizmann Institute of Science, Rehovot, Israel
		\textnormal{($^*$Contributed equally)}}
	\author{Ariel Nakav}
	\affiliation{Department of Physics of Complex Systems, Weizmann Institute of Science, Rehovot, Israel
		\textnormal{($^*$Contributed equally)}}
	\author{Lee Drori}
	\affiliation{Department of Physics of Complex Systems, Weizmann Institute of Science, Rehovot, Israel
		\textnormal{($^*$Contributed equally)}}
	\author{Ofer Firstenberg}
	\affiliation{Department of Physics of Complex Systems, Weizmann Institute of Science, Rehovot, Israel
		\textnormal{($^*$Contributed equally)}}

	\begin{abstract}
		Realizing strong interactions between individual photons, mediated by matter, is a cornerstone for advancing photonic quantum computing and quantum nonlinear optics. 
		In such systems, the interaction range is typically limited by the narrow bandwidth of the material excitation, resulting in a tradeoff between interaction strength and pulse duration.
		Here, we address this limitation by exploring interactions between counter-propagating photons mediated by Rydberg polaritons. We experimentally demonstrate strong photon-photon interactions, achieving a record-long anti-correlation range exceeding $1~\mus$.
		This extended range enables photon pulses that are long enough to fit within the polariton bandwidth, yet short enough to remain within the interaction range.
		Under these conditions, we observe complete photon blockade of entire pulses, tunable via their relative timing. 
		Extending to the three-photon regime, we observe enhanced interactions when a photon encounters a counter-propagating pair.  
		These results, supported by analytical theory and numerical simulations, establish counter-propagating Rydberg polaritons as a powerful platform for engineering interactions in quantum light fields, enabling deterministic photonic operations and opening new directions in few-photon quantum dynamics.
	\end{abstract}
	
	\maketitle

	\section{Introduction}
	Quantum nonlinear optics explores phenomena arising from strong effective photon-photon interactions, enabling the creation and manipulation of nonclassical light \cite{chang2014quantum,kala2025opportunities}. Unlike classical nonlinear optics, which requires intense fields, quantum nonlinear optics operates at the single-photon level. Realizing such interactions is essential for photonic quantum computing \cite{Gorniaczyk2014SinglePhoton,Tiarks2019NatPhys,Stolz2022PRX}, quantum state engineering \cite{chang2008crystallization,rosenblum2016extraction,Otterbach2013WignerCrystallization,Firstenberg2013AttractivePhotons,clark2020observation,cantu2020repulsive,stiesdal2021controlled}, and fundamental explorations of quantum few-body physics \cite{bienias2014scattering,Gullans2016EffectiveField,roy2017colloquium,jia2018strongly,das2025multiband}. 
	
	While the framework of linear-optics quantum computing  \cite{knill2001scheme,Kok2007}, based on interference of photons in beam splitters, has enabled remarkable progress,  it relies on inherently probabilistic operations \cite{nakav2025quantum,okamoto2005demonstration,kiesel2005linear}. Achieving deterministic photon-photon interactions therefore remains a central challenge for scalable photonic quantum technologies. Several platforms have been explored to realize strong optical nonlinearities at the single-photon level, including quantum emitters \cite{hoi2013giant,thompson2013coupling}, quantum dots \cite{Lodahl2015,faraon2008coherent}, cavity-based quantum networks \cite{Reiserer2015}, and trapped atoms \cite{Deutsch1997,birnbaum2005photon} or ions \cite{stute2012tunable}. 
	
	Among these, Rydberg polaritons have emerged as a leading approach \cite{Firstenberg2016NonlinearQuantum}. These polaritons---hybrid excitations of light and Rydberg atoms in ultracold ensembles---inherit their effective interaction from the long-range dipolar coupling between the Rydberg atoms. They enable hallmark phenomena such as photon blockade \cite{Peyronel2012QuantumNonlinear,Baur2013SinglePhoton}, few-photon bound states \cite{Firstenberg2013AttractivePhotons,Liang2012ObservationCoherent}, conditional phase shifts \cite{tiarks2016optical,thompson2017symmetry}, and quantum vortices of photons \cite{drori2023quantum,das2025multiband}.
	
	Many experiments on interactions between Rydberg polaritons have relied on co-propagating geometries \cite{Pritchard2009CooperativeAtomLight,Peyronel2012QuantumNonlinear}, where photons travel in the same direction, typically within a single continuous optical mode. While this configuration has enabled major advances, it imposes fundamental constraints on interaction range, symmetry, and tunability. Recent progress has strengthened the interactions by combining Rydberg media with optical cavities, enabling large and even deterministic nonlinearities \cite{vaneecloo2022intracavity,Tiarks2019NatPhys,tiarks2016optical}, albeit at the cost of increased experimental complexity and restrictions on the accessible optical modes. Alternative approaches employ stored Rydberg excitations interacting with propagating photons \cite{Gorniaczyk2014SinglePhoton,thompson2017symmetry,Baur2013SinglePhoton}, but are limited by storage efficiency and decoherence. These considerations motivate the exploration of traveling-wave configurations involving fully propagating photons. 
	
	In this work, we study counter-propagating photons. In contrast to co-propagating photons, counter-propagating photons do not need to enter the medium simultaneously in order to interact: in a sufficiently long medium, a photon entering early from one side can still overlap with a later photon entering from the opposite side, leading to an extended interaction time set by the propagation delay through the medium.
	Additionally, the interaction defines a richer regime of quantum nonlinear optics, as it couples distinct spatial modes. This opens new possibilities for two-qubit photonic gates, allowing for deterministic control over interaction strength and enabling few-body quantum effects inaccessible in co-propagating systems. 
	
	Theoretical studies have long predicted that counter-propagating photons in Rydberg media should give rise to distinct quantum nonlinear phenomena. 
	Early proposals by Friedler \textit{et al.} \cite{Friedler2005LongRangeInteractions} and Gorshkov \textit{et al.} \cite{Gorshkov2011PhotonPhoton} analyzed such geometries for implementing photonic quantum gates, showing that counter-propagating configurations can be more robust and less resource-intensive than their co-propagating counterparts. In these models, each photon effectively interacts with the full extent of the other, suppressing complex spatiotemporal correlations and enabling uniform nonlinear effects across the wavepacket. 
	Subsequent works examined the time-dependent dynamics of finite-size counter-propagating pulses, highlighting the roles of limited transmission bandwidth and interaction-induced dispersion \cite{Yang:16,Bing2014,Bienias_2020}. 
	Yet despite this promise, strong interactions between counter-propagating photons have remained largely unexplored experimentally.
	
	Here, we present the first experimental investigation of photon-photon interactions with unconfined counter-propagating photons. 
	We focus on the dissipative regime, where all photons couple to the same Rydberg state, and the interatomic van der Waals interaction leads to photon blockade: each photon suppresses the transmission of its counter-propagating partner, producing strong quantum anti-correlations. Operating first with continuous-wave light, we observe a strikingly long anti-correlation duration exceeding $1~\mus$, setting a record for the temporal range of photon-photon interactions. This range is more than twice that of co-propagating photons in our system, which itself constitutes a record for a single mode. 
	
	Motivated by this extended interaction range, we next study pulsed photon interactions. Finite-duration pulses are essential for quantum logic operations, which require deterministic interactions. For such operations to be efficient, the pulses must be long enough to remain within the system's transmission bandwidth, yet short enough to fully overlap within the interaction range. We identify an optimal pulse duration that satisfies both conditions, such that the bandwidth-induced attenuation and blockade infidelity each remain below 10\%.
	
	Finally, we explore three-photon interactions. While previous three-polariton experiments were limited to co-propagating geometries \cite{Liang2012ObservationCoherent,Porto2021,drori2023quantum}, we investigate the interaction of a single photon with a counter-propagating photon pair. Remarkably, while the pair exhibits dissipative self-interaction, the addition of a third photon enhances the suppression. 
	These results are supported by analytical and numerical models that describe counter-propagating interacting photons in nonuniform media, both in steady state and in the time-dependent regime. 
	
		\section{Experimental system}
	
\begin{figure}
		\center
		\includegraphics[width=.38\textwidth,trim={0cm 1.8cm 0cm 1.5cm},clip]{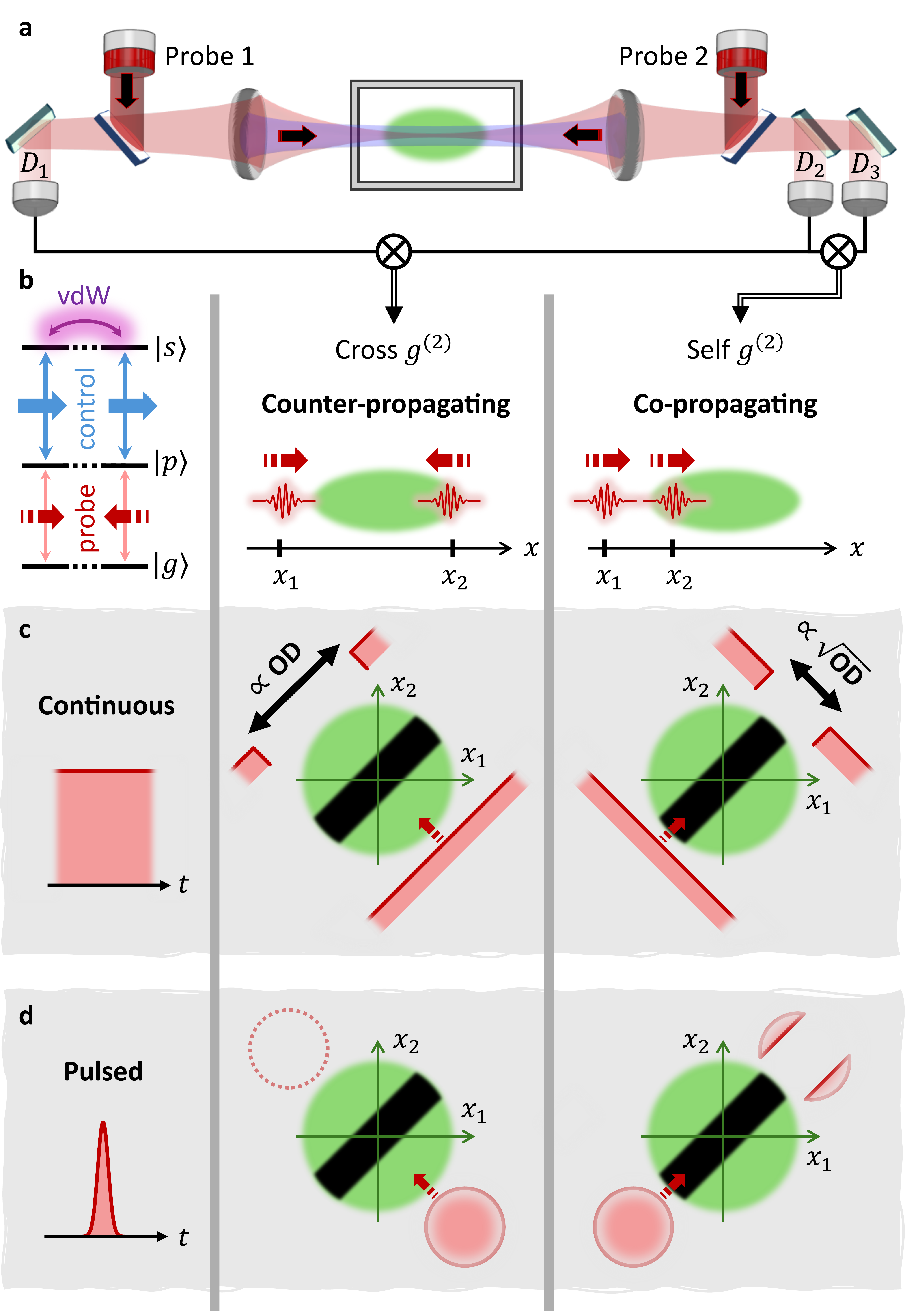}
		\caption{\textbf{Setup and two-photon dynamics.}
			\textbf{a}, Counter-propagating probe fields are sent from opposite directions into an elongated atomic cloud in the presence of a strong control field.
			Single-photon detectors $D_{1,2,3}$ measure correlations between counter-propagating photons (cross $\gtwo$ of Probe 1 and Probe 2) and between co-propagating photons (self $\gtwo$ of Probe 1).
			\textbf{b}, Level structure of $^{87}$Rb used to generate Rydberg polaritons interacting via van der Waals (vdW) coupling. The probe fields at 780 nm and control field at 480 nm couple the ground state $\ket{g}=5S_{1/2}$ to a high-lying Rydberg state $\ket{s}=90S_{1/2}$. Dissipative photon-photon interactions arise when operating on resonance with the short-lived intermediate state $\ket{p}=5P_{3/2}$, such that the Rydberg blockade leads to photon scattering. 
			\textbf{c}, Illustration of two-photon propagation dynamics in the ($x_1,x_2$) coordinate space, where $x_{1,2}$ denote the photons positions along the medium. The Gaussian atomic cloud appears as a circular region (green). Co-propagating photons traverse the ($x_1,x_2$)-space approximately parallel to the diagonal $x_1=x_2$, while counter-propagating photons traverse along the anti-diagonal $x_1=-x_2$. 
			Photons are scattered when $|x_2-x_1|<\rb$ due to the finite Rydberg blockade radius $\rb$, defining a diagonal interaction band (black). As a result, co-propagating photons interact only if they enter the medium simultaneously, whereas counter-propagating photons are guaranteed to cross the interaction band whenever both photons are inside the medium, even if they enter at different times. This leads to a markedly extended interaction range for counter-propagating photons,
			set by the medium length, with an anti-correlation time that scales linearly with the optical depth (OD). In contrast, co-propagating photons exhibit a shorter interaction range, limited by the blockade radius and broadened by the finite transmission bandwidth, scaling as $\sqrt{\OD}$.
			\textbf{d,} The extended interaction range for counter-propagating photons enables efficient deterministic operations with finite-duration pulses, provided the bandwidth-limited pulses fit entirely within the medium.
		}\label{fig:Figure 1}
	\end{figure}

	Our experiment begins by trapping and compressing an ultracold cloud of rubidium atoms elongated along the optical axis $x$, as shown in Fig.~\ref{fig:Figure 1}a. 
	To generate interacting Rydberg polaritons, we launch two probe beams from opposite ends of the cloud, along with a single control beam. The light fields couple the atomic ground state $|\mathrm{g}\rangle$ to a high-lying Rydberg state $|\mathrm{s}\rangle$ via an intermediate state $|\mathrm{p}\rangle$. Under electromagnetically induced transparency (EIT) conditions, the probe photons hybridize with Rydberg excitations to form slow-light polaritons that propagate through the medium.
	The corresponding level diagram is shown in Fig.~\ref{fig:Figure 1}b.
	
	Tuning all fields to resonance realizes an effective dissipative interaction between the polaritons \cite{Gorshkov2011PhotonPhoton}. When two polaritons approach within the Rydberg blockade radius $\rb\approx 10~\mum$, the strong van der Waals interaction shifts the Rydberg levels beyond the EIT linewidth, $2\gammaEIT= 10\cdot 2\pi$ MHz, 
	locally breaking the three-level transparency condition.
	One of the photons then experiences a resonant two-level response and is scattered. 
	The scattering probability is governed by the optical depth across the blockade radius, $\ODb$. 
	When $\ODb\gg 1$, this dissipative interaction leads to efficient photon blockade, manifesting as strong anti-correlations in the transmitted light.
	
	In our setup, the atomic cloud maintains a Gaussian density profile $\rho(x)\propto \exp{(-\pi x^2/L^2)}$, 
	centered at $x=0$, with a constant effective length $L\approx75~\mum$. 
	Over each 0.8-second experimental cycle, the density $\rho(x)$ decays slowly.
	In the continuous-wave experiments, the peak atomic density $\rho(0)= 3.3\times10^{12}~\mathrm{cm}^{-3}$ yields a total optical depth $\OD=\int \sigma_\mathrm{a}\rho(x)dx=72$, where the atomic absorption cross-section is
	$\sigma_\mathrm{a}=2.9\times 10^{-9}$~cm$^2$. 
	The corresponding peak blockade optical depth is given by $\ODb=(\rb/L)\OD= 10.4$ for $\rb= 10.85~\mum$, satisfying the strong-interaction condition.
	In the pulse experiments, $\OD=88$, $\rb= 9.95~\mum$, and $\ODb= 11.7$.
	
	Photon detection is performed at both ends of the medium 
	(see Fig.~\ref{fig:Figure 1}a). From the recorded detection times, we compute the second-order correlation function $\gtwo$ for both counter-propagating (cross) and co-propagating (self) photon pairs, and the third-order correlation function $\gthree$. These measurements allow us to quantify the effectiveness and temporal extent of the photon blockade under both continuous-wave and pulsed excitation.
	
	\subsection*{Technical details}

	The baseline setup and experimental procedure are described in detail in Ref.~\cite{drori2023quantum}. 
	We operate a time-modulated optical dipole trap, with probing performed during the 10-$\mus$-long dark windows of the modulation cycle.  
	Probe pulses are shaped using an arbitrary waveform generator driving an acousto-optic modulator. 
	To explore the continuous-wave regime, we use 9-$\mus$-long square pulses that approximate steady-state conditions with a photon rate of 0.1/$\mus$. To study finite-pulse interactions, we generate nearly Gaussian temporal profiles with full width at half maximum in the range $0.4\le T_\mathrm{w} \le 2.9~\mus$ and 0.5 photons per pulse. Each 0.8-second experimental cycle includes 40,000 dark windows, during which the optical depth gradually decays from 90 to 2.
	
	We generate the Rydberg polaritons via EIT in a ladder-level configuration with circularly-polarized light involving the ground state $|\mathrm{g}\rangle=|5S_{1/2}, F=2, m_F=2\rangle$, the intermediate state $|\mathrm{p}\rangle=|5P_{3/2},F=3,m_F=3\rangle$, and the Rydberg state $|\mathrm{s}\rangle=|90S_{1/2},J=1/2,m_J=1/2\rangle$. The van der Waals interaction coefficient for two $90S$ atoms is $C_6/\hbar = -16502\cdot 2\pi~\text{GHz}\cdot \mu\text{m}^6$.
	The probe fields at 780 nm are launched from opposite directions along the optical axis and are focused to a waist radius of $3.5~\mum$ in the medium. The co-axial control field at 480 nm enters from one side and is combined with, and later separated from, the probe beams using dichroic mirrors.
	
	Probe photons exiting the medium are detected by single-photon counting modules (SPCMs; Excelitas 780-14-FC). One SPCM is placed on one side of the medium and two on the opposite side, and the photon arrival times $t_{i=1,2,3}$ are recorded using a dedicated time tagger (Swabian Instruments Ultra).
	We compute $\gtwo_\mathrm{cross}(t_1,t_2)$, $\gtwo_\mathrm{self}(t_2,t_3)$, and $\gthree(t_1,t_2,t_3)$ using standard analysis procedures \cite{Peyronel2012QuantumNonlinear}, from which we extract averaged correlation functions, such as $\gtwo_\mathrm{cross}(\tau=t_1-t_2)$.
	
	\section{Interacting counter-propagating photons}\label{sec:cw}
	The dynamics of two interacting photons in an elongated medium can be intuitively visualized in the coordinate space ($x_1,x_2$), where $x_1$ and $x_2$ denote the positions of the two photons along the propagation axis $x$ (see Fig.~\ref{fig:Figure 1}c). In this picture, the Rydberg blockade defines a diagonal interaction band $|x_2-x_1|<\rb$, within which simultaneous Rydberg excitations are suppressed and photons are scattered. Efficient dissipative interaction thus requires a large optical depth across this band, quantified by $2\ODb$.

	A key ingredient in this dynamics is the finite transmission bandwidth of the EIT medium. 
	For a single polariton, the EIT resonance defines a spectral window of width $B=\gammaEIT/\sqrt{2\OD}$, such that frequency components outside this window are attenuated. As a result, the medium cannot support sharp temporal features on timescales shorter than $B^{-1}$, and short pulses broaden during propagation. 
	For two polaritons, this bandwidth limitation translates into a diffusion-like broadening of features in the relative coordinate $r=x_2-x_1$ \cite{Peyronel2012QuantumNonlinear}. In particular,  narrow depletion regions in the two-photon wavefunction, namely those created by the Rydberg blockade, smear and widen during propagation.

\begin{figure*}
		\centering\includegraphics[width=1.0\textwidth,clip]{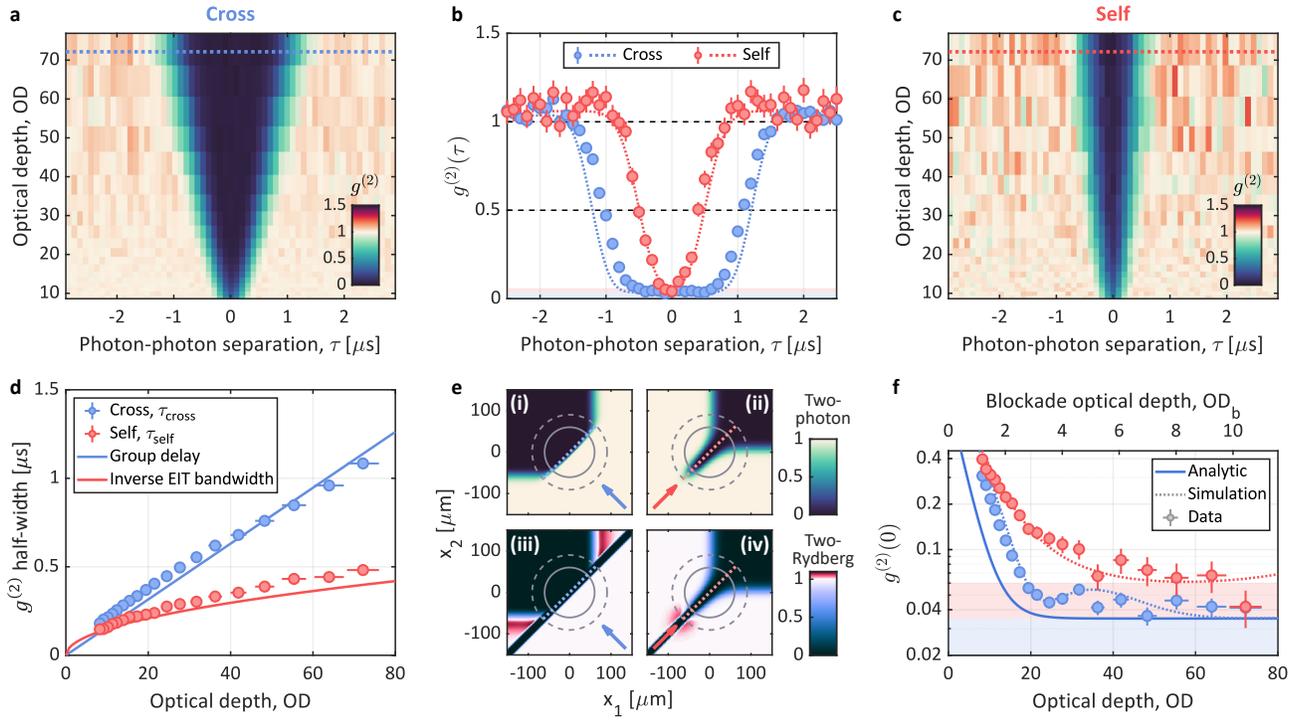}
		\caption{\textbf{Photon blockade in counter- and co-propagating configurations.\;}
			\textbf{a} and \textbf{c}, Normalized second-order correlation $\gtwo(\tau)$ as a function of the two-photon time separation $\tau$ measured for counter-propagating (`cross') and co-propagating (`self') continuous-wave probe fields, respectively, at various optical depths (OD). A pronounced quantum anti-correlation feature emerges in both cases, but broadens significantly with OD only in the cross configuration. 
			\textbf{b}, Direct comparison of cross and self $\gtwo(\tau)$ at $\OD=72$, highlighting the extended anti-correlation time of counter-propagating photons, $\tcross =1.08(1)~\mus$, compared to $\tself =0.48(2)~\mus$. Both represent record anti-correlation durations for initially independent photons.
			\textbf{d}, Extracted anti-correlation times.
			The blue line shows the group delay prediction for counter-propagating photons, $\tcross (\OD)=\OD/(2\gammaEIT)$, where $2\gammaEIT=10\cdot 2\pi$~MHz is the EIT linewidth. The red line shows the minimal-model prediction of inverse-bandwidth scaling for co-propagating photons, $\tself (\OD)=1.05/B\approx1.5\sqrt{\OD}/\gammaEIT$.
			\textbf{e}, Numerical simulations of the normalized two-photon (\textbf{i, ii}) and two-Rydberg (\textbf{iii, iv}) probability distributions in $(x_1,x_2)$-space, for counter-propagating (\textbf{i, iii}) and co-propagating (\textbf{ii, iv}) inputs. The atomic cloud is centered at $x=0$; solid and dashed circles mark the $2\sigma$ and $3\sigma$ contours of the Gaussian density profile ($\sigma=30.2~\mum$, $\OD=72$), and interactions occur for $x_1=x_2$ (dotted diagonal).
			\textbf{f}, Measured same-time correlations $\gtwo(0)$ for counter-propagating (blue) and co-propagating (red) photons. The solid line shows the analytic prediction for counter-propagation, $e^{-2\ODb}+s_\mathrm{cross}$, which overestimates the suppression due to simplifications in the minimal model.
			Dotted lines in (\textbf{b, f}) show full numerical simulations without fit parameters.
			Blue and red shaded regions in (\textbf{b, f}) indicate floor levels set by spectator photons  ($s_\mathrm{cross}=3.5\%$ and $s_\mathrm{self}=6\%$; see text), which are also included in the theoretical curves.
			The weak non-monotonicity close to the floor level arises from finite-size effects, where residual spatial modulations of the two-photon wavefunction originating at the boundaries of the medium and blockade region influence the output correlations.
			Error bars in (\textbf{b, d, f}) reflect OD uncertainty and one standard deviation from photon-counting statistics.
		}
		\label{fig:Fig2}
	\end{figure*}

	In the co-propagating configuration (Fig.~\ref{fig:Figure 1}c, right), both photons travel in the same direction, and their joint trajectory in ($x_1,x_2$)-space runs parallel to the diagonal $x_1=x_2$. As a result, wavefunction regions near the diagonal experience blockade throughout propagation, while off-diagonal regions remain essentially non-interacting.
	This creates a narrow, depleted feature with an initial width set by the blockade radius $\rb$.
	However, due to the finite EIT bandwidth, this feature broadens along the medium  [see panels (ii,iv) in Fig.~\ref{fig:Fig2}e], resulting in an anti-correlation duration on the order of $B^{-1}$, which increases with OD, scaling as $\sqrt{\OD}$ \cite{Peyronel2012QuantumNonlinear}.
	
	In contrast, counter-propagating photons (Fig.~\ref{fig:Figure 1}c, left) follow anti-diagonal trajectories parallel to $x_1=-x_2$, intersecting the blockade band over a much greater range, $|x_2+x_1|\lesssim L$, spanning the full medium length. Heuristically, in a long medium, a photon entering late (early) from one side can still overlap with an early (late) photon entering from the opposite side. As a result, the expected anti-correlation time scales linearly with the total optical depth, OD.
	This geometric reasoning suggests a dramatic difference in anti-correlation times $\tself $ and $\tcross $ for co- and counter-propagating photons. 
	
	We now turn to the experimental observation of these effects, presented in Fig.~\ref{fig:Fig2}.
	Figure~\ref{fig:Fig2}b shows a direct comparison of $\gtwo(\tau)$ for self and cross correlations at $\OD=72$, where $\tau$ is the time difference between detected photons. 
	In both cases, we observe deep anti-correlations dips: $\gtwo_\mathrm{self}(0)=0.04(1)$ and $\gtwo_\mathrm{cross}(0)=0.041(5)$. Despite their similar depths, the anti-correlation times differ markedly: the self-interaction half-width is $\tself = 0.48(2)~\mus$, while the cross-correlation extends to $\tcross =1.08(1)~\mus$---more than twice as long. These represent record anti-correlation durations for interacting, initially independent photons. 
	
	Note that both $\gtwo_\mathrm{self}(0)$ and $\gtwo_\mathrm{cross}(0)$ do not vanish completely at high OD but instead saturate at a finite floor level 
	(also shown in Fig.~\ref{fig:Fig2}f). These residual signals 
	are not dominated by background light or dark counts but rather originate from a small sub-population of `spectator' probe photons that traverse the medium without experiencing strong interaction. These photons, likely arriving from non-Gaussian tails of the focused beams (due to optical aberrations) and from imperfect polarization, are essentially non-interacting, thus setting a lower bound on the measured correlation functions \cite{Peyronel2012QuantumNonlinear}.
	
	To explore the dependence on optical depth, we measure $\gtwo(\tau)$ over a range of OD values. Figures~\ref{fig:Fig2}a and \ref{fig:Fig2}c show the evolving anti-correlation profiles for the cross and self configurations, respectively. While both deepen with increasing OD, the dip width grows much more rapidly in the counter-propagating case.
	This behavior is quantified in Fig.~\ref{fig:Fig2}d: the self-interaction width scales as $\tself \sim\sqrt{\OD}$, consistent with EIT bandwidth broadening, while the cross-interaction width grows linearly with OD, reflecting the full group delay through the medium.
	\clearpage
	
	To account for these trends, we introduce a minimal theoretical model based on a stationary two-polariton wavefunction evolving in ($x_1,x_2$)-space under an effective Hamiltonian. This model captures light propagation in a three-level EIT medium with interatomic van der Waals interaction. The dynamics reduce to an equation for two components (see Appendix B for derivation), which for a uniform-density medium takes the form
	\begin{align}
		\label{eq:diraq3}
		&\hspace{-3pt} \left[\hspace{-2pt}\begin{pmatrix}        
			{\pm\partial_{x_1}}&0\\    
			0&{\partial_{x_2}}
		\end{pmatrix}
		\hspace{-3pt}+\hspace{-2pt}\alpha_1\hspace{-3pt}-\hspace{-2pt}\frac{\alpha_2}{2}\begin{pmatrix}
			1&1\\1&1
		\end{pmatrix}\hspace{-2pt}
		\right]
		\hspace{-3pt}
		\begin{pmatrix}ES(x_1,x_2) \\ SE(x_1,x_2) \end{pmatrix}       =0\:. 
	\end{align}
	Here, $ES(x_1,x_2)$ and $SE(x_1,x_2)$ denote wavefunction amplitudes with one propagating photon and one Rydberg excitation at the respective positions, and the sign distinguishes co-propagating ($+$) and counter-propagating ($-$) geometries. 
	The effective one-photon absorption coefficient is $\alpha_1=1/(2\la)+\gammaEIT/c$, where $\la=L/\OD$ is the attenuation length in the absence of EIT. The interaction-dependent absorption coefficient is 
	$\alpha_2=[1-\mathcal{V}(x_2-x_1)]/(2\la)$, where $\mathcal{V}(x_2-x_1)=\rb^6/(\rb^6-ir^6)$ is an effective (complex) interaction potential incorporating the Rydberg blockade radius $\rb=\sqrt[6]{C_6/(2\hbar \gammaEIT)}$.
	
	While this two-component equation can be solved numerically, further analytical insight is gained by keeping only the symmetric component of the wavefunction, $\psi(x_1,x_2)=[ES(R,r)+SE(R,r)]/2$, where $R=(x_1+x_2)/2$ is the center-of-mass coordinate, and $r=x_2-x_1$ is the relative coordinate. This approximation, which neglects the antisymmetric evolution, has little effect on the observed correlations 
	\cite{Peyronel2012QuantumNonlinear,Firstenberg2013AttractivePhotons, das2025multiband}. 
	In this limit, the two-photon amplitude adiabatically follows the symmetric component, such that $|\psi|^2$ effectively represents the two-photon probability distribution used to extract the correlations.

	We then find that the two-polariton wavefunction $\psi(R,r)$ satisfies diffusion-like equations:
	For co-propagating photons, the evolution is governed by
	\begin{equation}\label{eq:co-propgating_single}
		\dfrac{\partial}{\partial R}\psi= 4\la\dfrac{\partial^2}{\partial r^2}\psi - \frac{1}{\la} \mathcal{V}(r)\psi,
	\end{equation}
	as derived in Ref.~\cite{Peyronel2012QuantumNonlinear}. For counter-propagating photons, we obtain
	\begin{equation}\label{eq:counter-propgating_single}
		\dfrac{\partial}{\partial r}\psi= \frac{\la}{2}\dfrac{\partial^2}{\partial R^2}\psi - \frac{1}{2\la} \mathcal{V}(r)\psi.
	\end{equation}
	The effective potential $\mathcal{V}(r)$ introduces loss, depleting the wavefunction amplitude when both photons are within the blockade radius $\rb$, while the diffusion terms ($\propto\la$) arise from the finite EIT bandwidth and describe the spreading of two-photon correlations.
	
	In the co-propagating case, the dissipative interaction introduces a localized loss feature around $r\approx0$, which broadens as the photons propagate along $R$. This diffusion-like broadening scales as $\sqrt{L\la}$ and yields an interaction distance of $a_\mathrm{self}= 2.1\la\sqrt{2\OD}$, as found analytically from Eq.~(\ref{eq:co-propgating_single}) \cite{Peyronel2012QuantumNonlinear}. After the first photon exits the medium and is detected, the second photon propagates outwards with group velocity $\vg=2\la \gammaEIT$, giving an anti-correlation duration $\tself =a_\mathrm{self}/\vg=1.05\sqrt{2\OD}/\gammaEIT=1.05/B$. This scaling is plotted as the red line in Fig.~\ref{fig:Fig2}d.
	
	In contrast, for the counter-propagating case, $r$ plays the role of the propagation coordinate. The interaction acts as a sharp loss feature localized around $r\approx0$, which depletes the wavefunction across the entire medium length, $a_\mathrm{cross}=L$, making diffusion in $R$ negligible. In temporal terms, this yields an anti-correlation duration equal to the total group delay in the medium, $\tcross =L/\vg=\OD/(2\gammaEIT)$, shown as the blue line in Fig.~\ref{fig:Fig2}d. 
	The minimal model captures the distinct scaling behavior of the self and cross configurations without any fit parameters. It predicts---consistently with the measurements---that the ratio $\tcross /\tself \approx\sqrt{\OD/9}$ exceeds unity for $\OD>9$.
	
	\begin{figure*}[t!]
		\includegraphics[width=1\textwidth,clip]{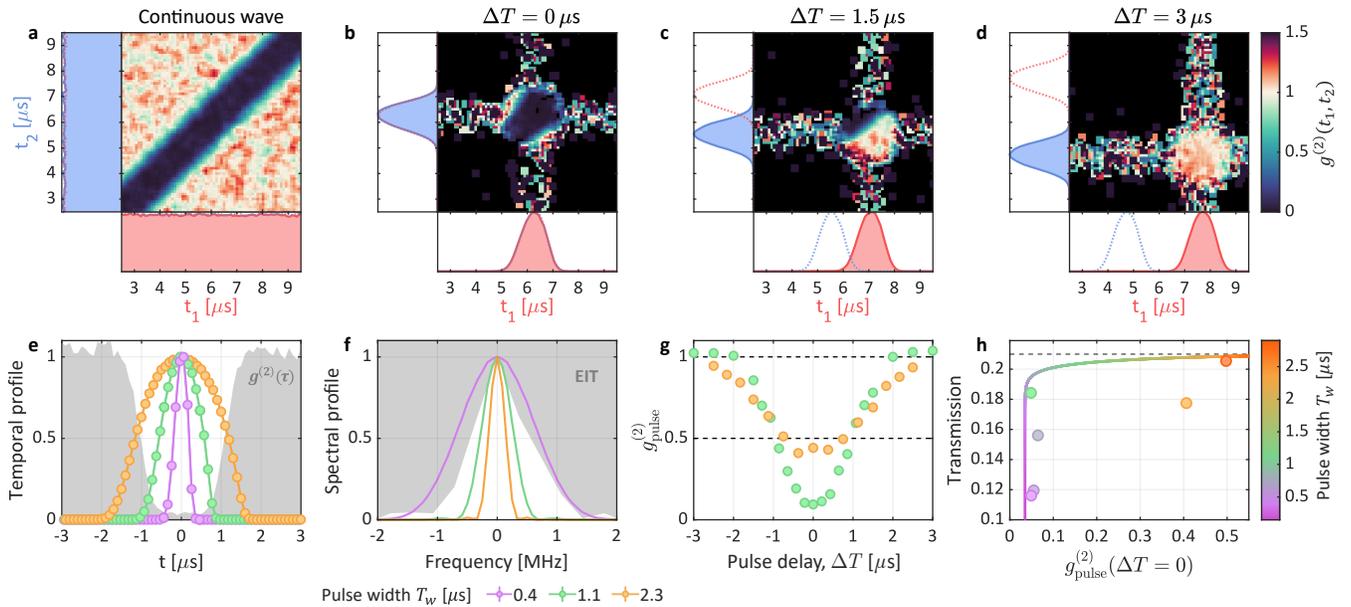}
		\caption{\textbf{Photon-photon interactions between counter-propagating pulses.}
			\textbf{a}-\textbf{d}, Measured two-photon cross-correlation functions $\gtwo(t_1,t_2)$ for continuous-wave (CW) input \textbf{(a)} and pulsed input with temporal delays $\Delta T = 0,~1.5,~3~\mus$ between counter-propagating 1.1-$\mus$-long pulses \textbf{(b-d)}.
			Side panels show photon arrival histograms, indicating the temporal profiles of the input pulses. OD = 72, $\gammaEIT=5\cdot 2\pi$ MHz (CW) and OD = 88, $\gammaEIT=8.5\cdot 2\pi$ MHz (pulsed).
			\textbf{e,} Temporal profiles of input pulses with full widths $T_\mathrm{w}=0.4$, $1.1$, and $2.3~\mu\text{s}$ (purple, green, yellow), overlaid with the anti-correlation range (non-shaded area) extracted from the CW $\gtwo(\tau)$ measurement. 
			\textbf{f,}  EIT transmission spectrum plotted alongside the spectral content of the input pulses. The 1.1-$\mus$ pulse fits within both the interaction range (\textbf{e}) and the EIT bandwidth (\textbf{f}).
			\textbf{g,} Pulse-level photon correlation $\gtwo_{\mathrm{pulse}}(\Delta T)$ as a function of temporal delay $\Delta T$, measured for 1.1- and 2.3-$\mus$ pulses. The coincident transmission of the 1.1-$\mus$ pulses is nearly completely suppressed.
			\textbf{h,} Pulse-level correlation $\gtwo_\mathrm{pulse}(\Delta T=0)$ and transmission $\mathcal{T}_\mathrm{pulse}$, measured for pulse widths $0.4\le T_\mathrm{w} \le 2.9~\mus$. The color-coded line shows the analytic predictions assuming Gaussian pulses:
			$\mathcal{T}_\mathrm{pulse}=\mathcal{T}_\mathrm{CW}/\sqrt{1+2\ln{2}/(B^2 T_\mathrm{w}^2)}$ and $\gtwo_\mathrm{pulse}(\Delta T=0)=1-\mathrm{erf}(\sqrt{2\ln{2}}\tcross /T_\mathrm{w})$ (see Appendix A)}
		\label{fig:pulses}
	\end{figure*}
	
	To visualize the photon dynamics inside the medium beyond the approximations of the minimal model, we perform full numerical simulations of the two-polariton wavefunction. These simulations incorporate the Gaussian atomic density profile and finite decoherence of the atomic excitations \cite{Peyronel2012QuantumNonlinear,drori2023quantum}. The resulting two-photon and two-Rydberg probabilities, $|EE(x_1,x_2)|^2$ and $|SS(x_1,x_2)|^2$ (normalized to the non-interacting case), are shown in Fig.~\ref{fig:Fig2}e. In both geometries, the strong blockade suppresses Rydberg pair excitations near the diagonal $x_1\approx x_2$, as evident in the $SS(x_1,x_2)$ maps (panels iii,iv). For the co-propagating photons, the amplitude $EE(x_1,x_2)$ exhibits localized depletion near the diagonal, which broadens along the medium due to finite bandwidth (ii). This broadening corresponds to the observed anti-correlation width $\tau_\mathrm{self}$.
	In contrast, for counter-propagating photons (i,iii), the depletion at the diagonal is sharp and spans a much larger spatial region. The extended suppression of $EE(x_1,x_2)$ reflects the longer interaction time, consistent with the observed scaling of $\tau_\mathrm{cross}$.
	
	Figure~\ref{fig:Fig2}f shows the measured same-time two-photon correlation $\gtwo(0)$ as a function of OD and $\ODb$ for both geometries. 
	In the co-propagating case, bandwidth-induced broadening delays the suppression of $\gtwo_\mathrm{self}(0)$, whereas in the counter-propagating case, $\gtwo_\mathrm{cross}(0)$ decreases more rapidly with OD. An analytical expression for $\gtwo_\mathrm{cross}(0)$ can be derived from Eq.~(\ref{eq:counter-propgating_single}) by neglecting diffusion and solving for $\psi(R=0,r)$ over $r\in[-L,L]$. In the limit $L\gg\rb$, this yields $\gtwo_\mathrm{cross}(0)=|\psi(0,L)/\psi(0,-L)|^2\approx e^{-2\ODb}$, 
	consistent with the geometric picture of exponential attenuation across the blockade region. This prediction (solid line in Fig.~\ref{fig:Fig2}f) overestimates the suppression and deviates from the full numerical simulations (dotted line), which closely match the experimental data. The discrepancy arises from two simplifying assumptions in the minimal model (with comparable contributions): uniform density and adiabatic elimination of intermediate states, the latter failing to capture the abrupt entrance and exit from the blockade region along $r$.
	
\section{Interactions between photon pulses}\label{sec:pulses}

Deterministic quantum logic operations require strong interactions between finite-time photonic pulses, with the entire wavepacket of one photon interacting with the other during their traversal through the medium. A key limitation of the co-propagating configuration is that the interaction range $\tself \approx1/B$ scales inversely with the EIT transmission bandwidth $B$, which by definition also sets the minimal pulse width that can propagate without significant loss. As a result, pulses short enough to interact effectively are too spectrally broad to transmit, while longer, bandwidth-limited pulses transmit well but experience only partial blockade.

Counter-propagating photons overcome this constraint at high $\OD$. While their minimal pulse width remains limited by the bandwidth, their interaction range, $\tcross \approx\sqrt{\OD}/(3B)$, grows significantly for $\OD\gg 9$. This separation of scales enables pulse widths that fit simultaneously within both the transmission windows and the interaction range, thus enabling deterministic interactions between entire photon pulses.

We test this directly by sending pairs of counter-propagating pulses through the medium and measuring their normalized two-photon correlation function $\gtwo(t_1, t_2)$, where $t_1$ and $t_2$ are the detection times of the outgoing photons. A pulse with full width of $T_\mathrm{w}=1.1~\mus$ is chosen to satisfy both the interaction range and bandwidth requirements, and we control the relative delay $\Delta T$ between the counter-propagating pulses. Figure~\ref{fig:pulses}a shows the continuous-wave (CW) baseline, highlighting a diagonal depletion band of width $\sim\sqrt{2}\tcross $, while Figs.~\ref{fig:pulses}b-d present results for pulses with increasing relative delays. Since depletion occurs only near the $\gtwo(t_1, t_2)$ diagonal, unsynchronized pulses ($\Delta T=1.5,~3~\mus$, Figs.~\ref{fig:pulses}c,d)  experience only partial suppression. In contrast, synchronized pulses ($\Delta T=0$, Fig.~\ref{fig:pulses}b) show near-complete extinction of coincident transmission, indicating effective and deterministic interaction. We note that the pulse durations are only weakly modified during propagation ($\lesssim10\%$ broadening at the highest OD), and this is accounted for in the analysis of the correlations.

A subtle V-shaped broadening appears along the diagonal in the pulsed $\gtwo(t_1,t_2)$ map (Fig.~\ref{fig:pulses}b), reflecting spatiotemporal dispersion absent in the CW case \cite{Yang:16,Bienias_2020}. This effect arises from the dynamics of the finite-duration two-photon wavefunction and is captured by a minimal time-dependent model presented in Appendix B.

To identify the optimal pulse width, we compare temporal profiles and spectral content of several input pulses to the constraints imposed by the medium.
Figure~\ref{fig:pulses}e displays pulses of widths $T_\mathrm{w}=0.4, 1.1, 2.3~\mus$, overlaid on the anti-correlation curve from the CW data. While the 0.4 and 1.1-$\mus$ pulses fit within the interaction range, the 2.3-$\mus$ pulse exceeds it and is expected to experience incomplete blockade. Figure~\ref{fig:pulses}f presents the same pulses in the frequency domain, alongside the EIT transmission window. Only the 1.1 and 2.3-$\mus$ pulses lie entirely within the bandwidth, while the 0.4-$\mus$ pulse is too spectrally broad to transmit efficiently. Together, these constraints point to an optimal pulse width around 1 $\mus$.

To quantify the interaction strength between pulses, we define a pulse-level correlation metric, $\gtwo_\mathrm{pulse}(\Delta T)$. Here, $\Delta T$ is the relative timing between the centers of the two input pulses, and $\gtwo_\mathrm{pulse}(\Delta T)$ is calculated by correlating photon detections over the full pulse duration, \textit{i.e.}, using a single correlation time-bin of width $2T_\mathrm{w}$. As shown in Fig.~\ref{fig:pulses}g, synchronized 1.1-$\mus$ pulses exhibit near-complete suppression, with $\gtwo_\mathrm{pulse}(\Delta T=0)\ll1$, while the 2.3-$\mus$ pulses show only partial blockade. In both cases, the anti-correlation vanishes with pulse separation $\Delta T$, confirming that the interaction can be tuned via timing.

The pulse-level transmission and correlation are shown in Fig.~\ref{fig:pulses}h for different pulse durations $T_\mathrm{w}$. As expected, the transmission increases with $T_\mathrm{w}$ and asymptotically approaches the CW limit. In contrast, the anti-correlation $\gtwo_\mathrm{pulse}(\Delta T{=}0)$ deepens for shorter pulses, approaching the CW value for coincident photons, $\gtwo_\mathrm{cross}(\tau{=}0)$. The theoretical curve represents analytical predictions for Gaussian pulses and captures the observed trade-off between transmission and interaction strength (see Appendix A). 
This mapping defines a practical design space for optimizing pulse-based quantum logic in dissipative photon-photon interaction regimes.

	\begin{figure*}[!htb]
	\begin{minipage}[c]{0.48\textwidth}
		\noindent\justifying
		\includegraphics[width=.9\columnwidth]{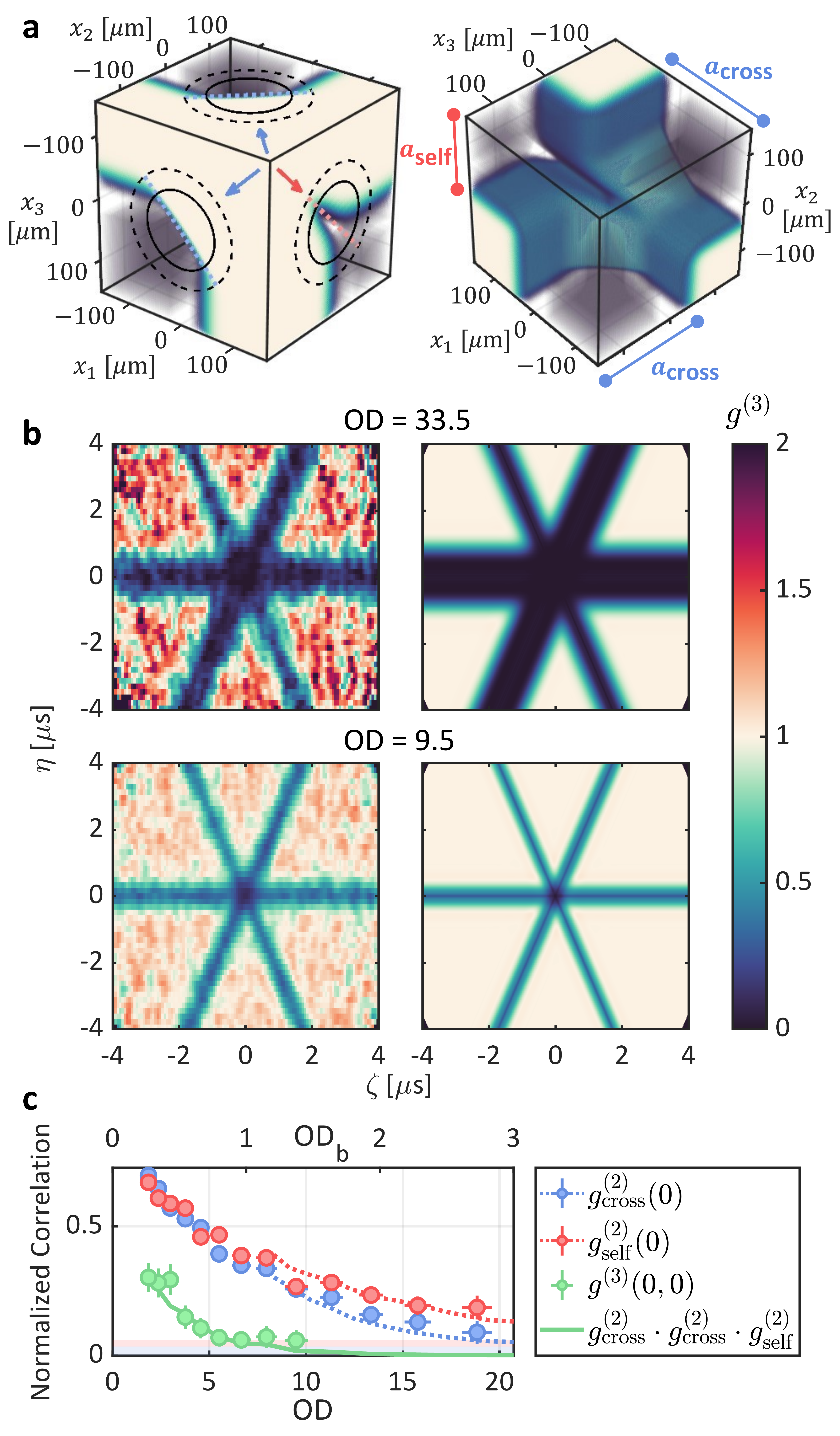}
	\end{minipage}
	\vspace{0pt}
	\hfill
	\begin{minipage}[c]{0.48\textwidth}
		\vspace{-20pt}
		\noindent
		\justifying\caption{\textbf{Three-photon interactions in counter-propagating geometry.}
			\textbf{a}, Calculated three-photon wavefunction $\psi(x_1,x_2,x_3)$, where photon 1 counter-propagates with photons 2 and 3, for $\OD=33.5$. Cross-interactions occur for $x_1=x_2$ and $x_1=x_3$ (blue dotted lines), while self-interaction occurs along $x_2=x_3$ (red dotted line). The Gaussian cloud boundary is marked by its projections (circles).  The entrance view (left cube) shows the incoming uncorrelated wavefunction (beige, propagation directions indicated by arrows) developing anti-correlations (transparent blue) inside the medium. The exit view (right cube) reveals the distinct spatial ranges of the cross- and self-interactions, $a_\mathrm{cross}$ and $a_\mathrm{self}$. 
			\textbf{b}, Measured three-photon correlation function $\gthree(\eta,\zeta)$ (left) and corresponding simulations (right), where $\eta$ and $\zeta$ are Jacobi coordinates representing time separations between photon detections. At high OD (top; $\OD=33.5$), the three-photon suppression $\gthree(0,0)\ll 1$ is already saturated by two-photon blockade, $\gtwo_\mathrm{cross}(0)\ll 1$. At low OD (bottom; $\OD=9.5$), while the two-photon blockade is incomplete $\gtwo_\mathrm{cross}(0)=0.26$, the three-photon suppression remains strong, reaching the floor level $\gthree(0,0)=0.06$.
			\textbf{c}, Measured three-photon coincidence $\gthree(0,0)$ as a function of $\OD$ and $\ODb$ (green circles), compared to two-photon coincidence data (blue and red), highlighting the crossover near $\ODb\approx 3$ between the pairwise three-photon and the saturated two-photon interaction regimes. For $\gtwo_\mathrm{cross}(0)$ and $\gtwo_\mathrm{self}(0)$, we show values from direct two-photon correlations (dashed lines) and from off-centered regions in $\gthree(\eta,\zeta)$ (circles). The naive expectation from independent pairwise blockade, $[\gtwo_\mathrm{cross}(0)]^2\times\gtwo_\mathrm{self}(0)$ (solid line), agrees well with the observed three-photon suppression. Shaded regions indicate floor levels $s_\mathrm{cross}$ and $s_\mathrm{self}$ due to spectator photons.
		}
		\label{fig:three_photons}
	\end{minipage}
\end{figure*}

\section{Three-photon interactions}\label{sec:three_photons}

While two-photon blockade produces strong anti-correlations, it does not fully capture the complexity of photon-photon interactions in quantum nonlinear media. In particular, three-photon dynamics reveal few-body effects that cannot generally be reduced to independent pairwise processes---especially when a single photon can block the interaction between multiple others. This mechanism is known to saturate the conditional phase shift in the dispersive, co-propagating regime \cite{drori2023quantum}.  Here, we explore three-photon interactions in a counter-propagating geometry, where a single photon collides with a photon pair. This configuration enables simultaneous cross- and self-interaction within the same system, providing direct access to their interplay.

Figure~\ref{fig:three_photons}a shows a simulated three-photon wavefunction $\psi(x_1,x_2,x_3)$, where photon 1 counter-propagates with photons 2 and 3 in a Gaussian atomic medium. In this geometry, interactions occur simultaneously along two cross channels ($x_1=x_2$, $x_1=x_3$) and one self channel ($x_2=x_3$). We display both the input faces in the $(x_1,x_2,x_3)$-space, viewed from the $(x,-x,-x)$ corner, and the output faces, viewed from $(-x,x,x)$. The initially uncorrelated wavefunction exhibits strong suppression along all interaction diagonals. On the input faces, where one photon is still outside the medium, $\psi(x_1,x_2,x_3)$ reduces to pairwise interactions: the $x_2x_3$ face reproduces the behavior of co-propagating photons (cf.~Fig.~\ref{fig:Fig2}e.ii), while the $x_1x_2$ and $x_1x_3$ faces manifest the counter-propagating behavior (Fig.~\ref{fig:Fig2}e.i). 
The extent of the depleted regions reflects the different spatial ranges of cross and self interactions. 
While these results are obtained numerically, we provide in Appendix B a minimal model for three interacting photons in counter-propagating geometry. 

We measure the three-photon correlation function $\gthree(\eta,\zeta)$ of the transmitted light over a range of optical depths, as demonstrated in Fig.~\ref{fig:three_photons}b (left). The relative photon detection times are expressed using Jacobi coordinates $\eta=(t_1-t_2)/\sqrt{2}$ and $\zeta=(t_1+t_2-2t_3)/\sqrt{6}$, where $t_1$ corresponds to a detection on the opposite side of the medium from $t_2$ and $t_3$. In this frame, $\gthree(0,0)$ corresponds to coincident triplets, while off-center regions reflect pairwise correlations, either $\gtwo_\mathrm{cross}$ or $\gtwo_\mathrm{self}$.

At high optical depth (Fig.~\ref{fig:three_photons}b, top), we observe strong anti-correlations $\gthree\ll 1$ along all pairwise interaction lines, indicating efficient blockade of both cross and self interactions. 
At low OD (bottom), although the pairwise blockade is weaker [$\gtwo(0)\approx 0.3$], the suppression of three-photon coincidences remains substantial [$\gthree(0,0)=0.06(4)$]. This enhanced three-photon suppression, arising from the combined effect of two cross- and one self-interaction, is summarized in Fig.~\ref{fig:three_photons}c. 

The different widths of the anti-correlation features in $\gthree(\eta,\zeta)$ reflect the interaction range ratio $\tcross /\tself \approx\sqrt{\OD/9}$. All observed trends are reproduced by full numerical simulations (Fig.~\ref{fig:three_photons}b, right), which account for the time-dependent outward propagation following the first and second photon detection events \cite{das2025multiband}.

\section{Discussion}

We have explored photon-photon interactions in a counter-propagating geometry. This configuration realizes a qualitatively distinct regime of quantum nonlinear optics, where the interaction range grows linearly with optical depth and exceeds the limit imposed by the EIT bandwidth. Operating in this regime, we demonstrated full-pulse photon blockade, where the entire photonic wavepackets interact deterministically.

This separation of scales allowed us to identify a pulse duration that fits within both the transmission window and the interaction range, paving the way for deterministic few-photon logic primitives. Going beyond pairwise effects, we investigated the dynamics of three interacting photons and observed enhanced suppression when a photon is blocked by a counter-propagating photon pair.

These results establish counter-propagating Rydberg polaritons as a powerful platform for studying few-photon physics in nonlinear media. The ability to reach and control deterministic interactions through timing opens new directions for optical quantum gates and few-body dynamics in strongly interacting systems. Looking beyond the dissipative regime, dispersive counter-propagating interactions realized via off-resonant EIT could enable coherent phase gates and topologically nontrivial quantum vortices, opening a rich landscape for quantum state engineering.

\appendix

\section*{Appendix A: Analytical estimates for pulse-level transmission and correlation}

To model the observed relationship between pulse transmission and two-photon correlations, we consider Gaussian probe pulses and an approximate analytical treatment of EIT filtering and counter-propagating photon blockade. We assume normalized input pulses with temporal intensity profile $I(t)=\sqrt{2/\pi} e^{-2t^2/(T_\sigma^2)}/T_\sigma$, where $T_\sigma=T_\mathrm{w}/\sqrt{2\ln{2}}$. The corresponding power spectral profile is $\tilde{I}(\omega)=T_\sigma/\sqrt{2\pi} e^{-T_\sigma^2\omega^2/2}$.  

The EIT transmission spectrum, derived from the complex linear susceptibility, is given by $\mathcal{T}(\omega)=e^{-\OD\cdot\mathrm{Re}[1+\gammaEIT/(\gamma+i\omega)]^{-1}}$, where $2\gammaEIT$ is the EIT linewidth and $\gamma$ is Rydberg-excitation decoherence rate. 
In the limit $\omega, \gamma \ll \gammaEIT$, we approximate the spectrum as a Gaussian, $\mathcal{T}(\omega)\approx \mathcal{T}_\mathrm{CW} e^{-\omega^2/(2B^2)}$, where $\mathcal{T}_\mathrm{CW}=e^{-\OD \gamma/\gammaEIT}$ is the EIT transmission for a CW probe. Integrating over the pulse spectrum, the pulse-level transmission becomes 
\begin{equation}
	\mathcal{T}_\mathrm{pulse}=\int\limits_{-\infty}^{\infty}  \mathcal{T}(\omega)\tilde{I}(\omega) d\omega=\frac{\mathcal{T}_\mathrm{CW}}{\sqrt{1+1/(B^2 T_\sigma^2)}}.
\end{equation}

To estimate the pulse-level correlation for synchronized pulses, $\gtwo_\mathrm{pulse}(0)$, we adopt a geometric picture of photon blockade in the temporal ($t_1,t_2$)-space, where the two pulses have the shape $I(t_1,t_2)=(2/\pi)e^{-2(t_1^2+t_2^2)/T_\sigma^2}/T_\sigma^2$. We assume an idealized anti-correlation region centered on $t_1=t_2$, corresponding to the dark strip in Figs.~\ref{fig:Fig2}a-d, within which coincident detection is fully suppressed and outside of which no correlation develops. The width of this strip, defined by $|t_1-t_2|<\tcross$, is $\sqrt{2}\tcross$. The blockade probability is therefore the overlap between the Gaussian pulse and the blocked region:
\begin{align}
	1-\gtwo_\mathrm{pulse}(0)&=\int\limits_{-\infty}^{\infty}\hspace{-5px}dt_1 \hspace{-7px}\int\limits_{-\frac{\tcross}{\sqrt{2}}}^{\frac{\tcross}{\sqrt{2}}}\hspace{-11px}dt_2\; I(t_1,t_2) =
	\mathrm{erf}\left(\frac{\tcross}{T_\sigma}\right)
\end{align}
where we employed the rotation symmetry of the pulse profile $I(t_1,t_2)$ in ($t_1,t_2$)-space.  
The expressions for $\mathcal{T}_\mathrm{pulse}$ and $\gtwo_\mathrm{pulse}(0)$ are used to generate the color-coded theoretical curve shown in Fig.~\ref{fig:pulses}h.

\vspace{0.5cm}

\section*{Appendix B: Minimal two- and three-polariton models}

To describe the evolution of counter-propagating Rydberg polaritons, we develop a minimal theoretical framework based on the stationary wavefunction of two polaritons in an effective EIT medium with interatomic van der Waals (vdW) interactions.
The dynamics of a single polariton in the medium are governed by a stationary Schrödinger-like equation, $H^{\pm}(x)\psi(x)=0$, where the effective Hamiltonian is given by
\begin{equation}\label{eq:H}
	H^{\pm}(x)= 	
	{-}\begin{pmatrix}
		\pm ic\partial_x &  \sqrt{\rho(x)}g & 0\\	
		\sqrt{\rho(x)}g   & \Delta+
		\rmi\Gamma &  \Omega \\
		0& \Omega & \delta+\rmi\gamma\\
	\end{pmatrix},
\end{equation}
with the sign $\pm$ corresponding to right- and left-propagating photons, and $c$ the speed of light. 
The polariton wavefunction $\psi(x)=[E(x),P(x),S(x)]^T$ consists of the amplitudes $E(x)$ for a single photon, $P(x)$ for a collective intermediate-state excitation, and $S(x)$ for a collective Rydberg excitation. The relevant atomic and optical parameters are: intermediate-state decay rate (half-linewidth) $\Gamma$; collective probe-photon coupling strength $\sqrt{\rho(x)}g$, where $g=\sqrt{c\Gamma\sigma_\mathrm{a}/2}$; control-field Rabi frequency $\Omega$; 
Rydberg-excitation decoherence rate 
$\gamma$; and the frequency detunings $\Delta$ and $\delta$. We consider the dissipative-interaction regime in which all fields are on resonance, \textit{i.e.}, $\Delta=\delta=0$, and neglect Rydberg decoherence by setting $\gamma=0$. 

To describe the interaction between two polaritons, we define a nine-amplitude wavefunction over the coordinates $(x_1,x_2)$ and introduce the vdW potential $C_6/|x_2-x_1|^6$, which acts on the Rydberg-Rydberg amplitude $SS(x_1,x_2)$. 
Assuming a sufficiently high atomic density to satisfy the slow-light condition $\rho g^2\gg \Omega^2$ (that is, $\vg\ll c$), we adiabatically eliminate the intermediate atomic state \cite{Peyronel2012QuantumNonlinear} and reduce the model to a dual-band propagation equation \cite{Gorshkov2011PhotonPhoton,das2025multiband}: 
\begin{equation}
	\label{eq:diraq2}
	\hspace{-7pt}
	\left[\hspace{-2pt}
	\begin{pmatrix}        
		\pm \partial_{x_1}&0\\    
		0&\partial_{x_2}
	\end{pmatrix}\hspace{-2.5pt}\mathrm{+}\frac{\Omega^2}{c\Gamma}\mathrm{+}\frac{g^2 }{2 c\Gamma}M\hspace{-1pt}(x_1,x_2)
	\hspace{-1.5pt}\right ]\hspace{-4pt}\begin{pmatrix}ES(x_1,x_2) \\ SE(x_1,x_2) \end{pmatrix} 
	\hspace{-2pt}=\hspace{-1.5pt}
	0\,, \hspace{-3pt}
	\vspace{0.5mm}
\end{equation}
where the $\pm$ sign distinguishes the co-propagating ($+$) and counter-propagating ($-$) geometries, and
\begin{equation}
	M=\begin{pmatrix}
		(\mathcal{V}+1)\rho(x_1)&(\mathcal{V}-1)\sqrt{\rho(x_1)\rho(x_2)}\\
		(\mathcal{V}-1)\sqrt{\rho(x_1)\rho(x_2)}&(\mathcal{V}+1)\rho(x_2)
	\end{pmatrix}.\nonumber
\end{equation}
The two components $ES(x_1,x_2)$ and $SE(x_1,x_2)$ represent wavefunction amplitudes with one propagating photon and one Rydberg excitation at the respective positions. These suffice to capture both the linear propagation and the two-body interaction dynamics under this approximation. The effective interaction potential $\mathcal{V}=\mathcal{V}(x_2-x_1)=\rb^6/(\rb^6-ir^6)$ incorporates the Rydberg blockade radius, defined as $\rb=\sqrt[6]{C_6/(2\hbar \gammaEIT)}$, where $\gammaEIT=\Omega^2/\Gamma$.
The two-photon wavefunction is obtained by numerically integrating Eq.~(\ref{eq:diraq2}) in the ($x_1,x_2$)-space.

To gain analytical insight into the dynamics, one can simplify the model by considering a uniform atomic density, for which the attenuation length $\la=c\Gamma/(2\rho g^2)=L/\OD$ is constant. In this case, Eq.~(\ref{eq:diraq2}) reduces to Eq.~(\ref{eq:diraq3}) in the main text.

For completeness, we provide here extensions of Eq.~(\ref{eq:diraq3}) to describe time-dependent two-photon dynamics and stationary three-photon dynamics. The first, relevant to the pulsed experiments discussed in Sec.~\ref{sec:pulses}, is a time-dependent propagation equation for two polaritons:
\begin{equation}
	\label{eq:twothree}
	-\frac{1}{c}\partial_t\Psi\hspace{-1.5pt}=\hspace{-2pt}\begin{pmatrix}        
		{\pm\partial_{x_1}}&0&0\\    
		0&{\partial_{x_2}}&0\\
		0&0&0
	\end{pmatrix}
	\hspace{-3pt}\Psi \hspace{-1pt}+\hspace{-2pt} \begin{pmatrix}        
		\alpha_1&0&\alpha_\mathrm{E}\\    
		0&\alpha_1&\alpha_\mathrm{E}\\
		\alpha_\mathrm{E}&\alpha_\mathrm{E}&2\alpha_\mathrm{E}^2/\alpha_2
	\end{pmatrix}
	\hspace{-3pt}\Psi\,,
\end{equation}
where $\Psi=[ES(x_1,x_2),SE(x_1,x_2), SS(x_1,x_2)]^\text{T}$ has three components, and $\alpha_\mathrm{E}=\sqrt{\gammaEIT/(2c\la)}$. While the stationary model [Eq.~(\ref{eq:counter-propgating_single})], captures the dominant anti-correlation range in the pulse experiments, it does not account for the weak broadening of $\tcross $ during the pulse time (manifested as the subtle V-shape dispersion along the diagonal in Fig.~\ref{fig:pulses}b). This spatiotemporal effect arises from the time-dependent interplay between Rydberg interactions and polariton dispersion, and can be reproduced numerically using Eq.~(\ref{eq:twothree}).   

The second extension, used to describe the stationary evolution of a single photon interacting with a counter-propagating pair, as studied in Sec.~\ref{sec:three_photons}, is given by
\begin{align}
	\label{eq:counter three}
	\hspace{-6pt}\left[  \hspace{-3pt}\begin{pmatrix}        
		{-\partial_{x_1}}&0&0\\    
		0&{\partial_{x_2}}&0\\
		0&0&{\partial_{x_3}}
	\end{pmatrix} \hspace{-3pt}
	{+}\alpha_1'{-}\frac{\alpha_3}{3} \hspace{-3pt}\begin{pmatrix}
		1&1&1\\
		1&1&1\\
		1&1&1
	\end{pmatrix} \hspace{-3pt}\right] \hspace{-3pt}
	\begin{pmatrix}ESS \\ SES \\ SSE \end{pmatrix} \hspace{-3pt}= \hspace{-2pt}0\,,
\end{align}
where the three wavefunction components $ESS(x_1,x_2,x_3)$, $SES(x_1,x_2,x_3)$, and $SSE(x_1,x_2,x_3)$ describe the propagation of one photon with two Rydberg excitations at the respective positions. The effective three-photon interaction is described by $\alpha_3=\left[2\la \left( 1+\frac{2 i}{3}\sum_{1\le i<j\le 3} r_\mathrm{b}^6/|x_i-x_j|^6\right)\right]^{-1}$, and $\alpha_1'=1/(2\la)+2\gammaEIT/c$.

\section*{Acknowledgments and Funding}
\noindent We acknowledge financial support from the Israel Science Foundation (3491/21, 1982/22), the US-Israel Binational Science Foundation and US National Science Foundation, the Minerva Foundation with funding from the Federal German Ministry for Education and Research, the Leona M.~and Harry B.~Helmsley Charitable Trust, and the Laboratory in Memory of Leon and Blacky Broder.

\section*{Disclosures}
\noindent The authors declare no conflicts of interest.
\bibliography{biblio}
\end{document}